\documentstyle[12pt]{article}
\renewcommand{\baselinestretch}{1.65}

\newcommand{\rA}{\rightarrow}

\begin{document}

\renewcommand{\baselinestretch}{1.1}

\title{Limits of space-times in five dimensions and their
                relation to the Segre Types}
\author{
F. M. Paiva\thanks{Departamento de Astrof\'{\i}sica, Observat\'orio Nacional, 
Rua General Jos\'e Cristino 77, 20921-400 Rio de Janeiro~--~RJ, Brazil, 
{\sc internet: fmpaiva@on.br}}
,\ \
M. J. Rebou\c{c}as\thanks{Centro Brasileiro de Pesquisas F\'{\i}sicas,
Rua Dr.\ Xavier Sigaud 150, 22290-180 Rio de Janeiro~--~RJ, Brazil, {\sc
internet: reboucas@cat.cbpf.br}}
 \ \ and \ \
A. F. F. Teixeira\thanks{Centro Brasileiro de Pesquisas F\'{\i}sicas,
Rua Dr.\ Xavier Sigaud 150, 22290-180 Rio de Janeiro~--~RJ, Brazil, {\sc
internet: teixeira@novell.cat.cbpf.br}}
}
\date{\today}
\maketitle
\begin{abstract}  \begin{sloppypar}
A limiting diagram for the Segre classification in 5-dimensional 
space-times is obtained, extending a recent work on limits of the
energy-momentum tensor in general relativity.
Some of Geroch's results on limits of space-times in general relativity 
are also extended to the context of five-dimensional 
Kaluza-Klein space-times.  \end{sloppypar}
\end{abstract}

{\sc pacs} numbers: 04.20.Cv \ \ 04.50.+h \ \ 04.20.Jb \ \ 04.20.-q 

\section{Introduction}
\label{Intro} 
\setcounter{equation}{0}

In general relativity, it is well known that the curvature tensor
can be uniquely decomposed into three irreducible parts, namely
the Weyl tensor, the traceless Ricci tensor and the Ricci scalar.
Petrov and others%
~\cite{Petrov1969,Penrose1960,PenroseRindler1986v2} 
have discussed the algebraic classification of the Weyl part, which
is known nowadays as Petrov classification.
The algebraic classification of the Ricci part, known as 
the Segre classification, has been discussed by several 
authors~\cite{Hall1984a} under different viewpoints, and is 
important, for example, in the characterization of matter distributions%
~\cite{HallNegm1986,Hall1984b,FerrandoMoralesPortilla1990,%
ReboucasTeixeira1991,SantosReboucasTeixeira1993},
as part of the procedure for checking local equivalence of
space-times%
~\cite{Cartan1951,Karlhede1980,MacCallumSkea1994,MacCallum1983,%
MacCallum1984,MacCallum1991}, 
and in the study of limits of non-vacuum solutions of Einstein's
field equations~\cite{PaivaReboucasMacCallum1993,%
PaivaRomero1993,PaivaReboucasHallMacCallum1996}.

In 1969, Geroch~\cite{Geroch1969} discussed some basic
properties of limits of space-times in general relativity (GR).
Regarding the algebraic types of the Weyl tensor he showed that
the Penrose specialization diagram~\cite{Penrose1960} 
for the Petrov classification is also a limiting diagram, that is to say,
under limiting processes only space-times with the same Petrov type
or one of its Penrose specializations can be reached. 

In a recent work Paiva {\em et al.\/}~\cite{PaivaReboucasHallMacCallum1996}
have investigated the relations among the Segre types of
the Ricci tensor under limiting processes in the framework of
general relativity. They have obtained a limiting diagram for 
the Segre classification of symmetric two-tensors in GR. 
Among the relevant consequences of their limiting scheme it is 
worth mentioning that it permits an extension of the coordinate-free 
approach to limits recently devised~\cite{PaivaReboucasMacCallum1993}
(see also~\cite{PaivaRomero1993}).

An essential result for the present article came out in a recent 
paper by Santos {\em et al.\/}~\cite{SantosReboucasTeixeira1995}
(see also Hall {\em et al.\/}~\cite{HallReboucasSantosTeixeira1996}), 
where they have performed the Segre classification of second order 
symmetric tensors on five-dimensional (5-D for short) Lorentzian spaces.

In the present paper we build a limiting diagram for the Segre
types in five-dimensional space-times, generalizing previous work on this 
matter~\cite{PaivaReboucasHallMacCallum1996}. 
We also extend some of Geroch's results on limits of space-times in 
general relativity to the context of 5-D Kaluza-Klein-type 
theories.
Although Paiva {\em et al.\/}~\cite{PaivaReboucasMacCallum1993}
coordinate-free procedure for finding out limits of space-times 
in GR has not yet been extended to 5-D space-times, the limiting 
diagram studied in the present work will certainly be relevant to 
any approach to possible limits of space-times in five dimensions.

Throughout this paper we shall use the concept of limit of a space-time
introduced in reference~\cite{PaivaReboucasMacCallum1993} (see also 
\cite{Geroch1969,Schmidt1987}), wherein 
by a limit of a space-time, broadly speaking, we mean a limit of a family 
of space-times when some free essential parameters are taken to a limit. 
So, for example, in the one-parameter family of Schwarzschild solutions
each member is a Schwarzschild space-time with a specific value 
for the mass parameter $m$. By space-time we understand a real 
differential manifold with a metric of Lorentzian signature
together with the attendant mathematical structures usually required in 
physics~\cite{HawkingEllis1973}. Finally we note that
although the Ricci tensor is  constantly referred to, the 
results of the following sections apply to any  second order real 
symmetric tensor defined on 5-dimensional Lorentzian spaces.

In the next section we present a brief summary of the main results 
on Segre classification in 5-D and discuss the minimal and
characteristic polynomial types corresponding to the Segre types in 5-D
space-times. These results are required for section~\ref{LimitDiag},
where we build a limiting 
diagram for the Segre type in 5-D. In the last section we discuss 
our main results and their extensions.

\section{Segre Types in 5-D }
\label{Segre5D} 
\setcounter{equation}{0}

The algebraic classification of the Ricci tensor in 5-D
space-times can be cast in terms of the eigenvalue problem
\begin{equation}
\label{eigen}
(R^{a}_{\ b} - \lambda \,\, \delta^{a}_{\ b}) \,V^{b}\, = 0,
\end{equation}
where $\lambda$ is a scalar, $V^{b}$ is a vector and the mixed
Ricci tensor $R^{a}_{\ b}$ is looked upon as a linear operator
$R: T_{p}(M) \longrightarrow T_{p}(M)$. Here and in what follows
$M$ is a real 5-dimensional space-time manifold locally endowed 
with a Lorentzian metric of signature $(+ - - - -), \; T_{p}(M)$ 
denotes the tangent space to $M$ at a point $p \in M$ and latin 
indices range from $0$ to $4$, unless otherwise stated.
Although the matrix $R^a_{\ b}$ is real, the eigenvalues $\lambda$
and the eigenvectors $V^b$ are often complex. A mathematical 
procedure used to classify matrices in such a case is to reduce 
them through similarity transformations to canonical forms 
over the complex field.
Among  the existing canonical forms the Jordan canonical form (JCF) 
turns out to be the most appropriate for a classification of 
$R^a_{\ b}$ in 5-D~\cite{SantosReboucasTeixeira1995}. In the Jordan
canonical form, a matrix consists of Jordan blocks along the
main diagonal. A Jordan block is, for example, a matrix of 
form
\vspace{2mm}
\begin{equation}
\left[ \begin{array}{cccc}
\lambda_1 &       1 &       0 &      0 \\
      0 & \lambda_1 &       1 &      0 \\
      0 &       0 & \lambda_1 &      1 \\
      0 &       0 &       0 & \lambda_1 
\end{array} \right] \,, \vspace{2mm} 
\end{equation}
where the equal elements along the main diagonal are 
the eigenvalue associated to the Jordan block. It is well known
that the Jordan canonical form is uniquely defined up to the
ordering of the Jordan blocks.

In the Jordan classification two square matrices are
said to be equivalent if similarity transformations exist 
such that they can be brought to the same JCF. The JCF of
a matrix gives explicitly its eigenvalues and makes apparent
the dimensions of the Jordan blocks.
However, for many purposes a somehow coarser classification
of a matrix is sufficient. In the Segre classification, 
for example, the value of the roots of the characteristic 
equation is not relevant --- only dimension of the Jordan 
blocks and degeneracy of eigenvalues matter.
The Segre type is a list $[n_1 n_2 \cdots n_r]$ of the 
dimensions of the Jordan blocks. Equal eigenvalues in distinct
blocks are indicated by enclosing the corresponding digits
inside round brackets. Thus, for example, in the degenerated Segre 
type $[(31)1]$ four out of the five eigenvalues are equal; 
there are three linearly independent eigenvectors, two of which
are associated to the Jordan blocks of dimensions 3 and 1, whereas
the last one corresponds to the block of dimension 1.

In classifying symmetric tensors in a Lorentzian space-time two 
refinements to the usual Segre notation are often used.
Instead of a digit to denote the dimension of a block with 
complex eigenvalue a letter is used, and the digit corresponding 
to a timelike eigenvector is separated from the others
by a comma.

As far as 5-D space-times are concerned, due to the Lorentz 
signature, the real eigenvectors of $R^a_{\ b}$ may be space-like, 
null or time-like. 
For these space-times, it can be shown~\cite{SantosReboucasTeixeira1995}
that some Segre types (as, for example, Segre types [5], [41], [32] and 
[221]) are not allowed because of the Lorentzian signature of the metric
and the symmetry of $R_{ab}$. One also learns from Santos {\em et al.\/}~%
\cite{SantosReboucasTeixeira1995} that the possible Segre types 
of $R^a_{\ b}$ in 5-D Lorentz spaces are:
\vspace{2mm}
\begin{enumerate}
\item   
 {[}1,1111] and its degeneracies  [1,11(11)], [(1,1)111], 
 [1,(11)(11)], \ [(1,1)(11)1], \  [1,1(111)], \  [(1,11)11], 
 [(1,1)(111)], \ [(1,11)(11)], \ [1,(1111)], \ [(1,111)1] \  and \ 
 [(1,1111)]. Type [(1,1111)] implies that $R_{ab}$ is
 proportional to the metric $g_{ab}$, it is usually referred to
 as Segre $0$.
\item
{[}2111]  and its specializations  [21(11)], [(21)11],  [(21)(11)],  
[2(111)], \ [(211)1] \  and \ [(2111)]. The first digit corresponds to
a null eigenvector while the others are associated to space-like
eigenvectors.
\item
 {[}311]  and its degeneracies  [3(11)], \  [(31)1] \  and \ [(311)].
Here again the first digit corresponds to a null eigenvector 
while the others correspond to space-like eigenvectors.
\item {[}$z\bar{z}$111] and its degeneracies [$z\bar{z}$(11)1] \
and \ [$z\bar{z}$(111)]. Here $z$ and $\bar{z}$ correspond to complex
conjugate eigenvectors with complex conjugate eigenvalues. The digits
correspond to space-like eigenvectors with real eigenvalues. 
\end{enumerate}

We shall now discuss the characteristic and minimal polynomials
in connection with Segre types and build a table, which will be
important in the derivation of the limiting diagram for the 
Segre classification of the next section.

Associated to the eigenvalue problem~(\ref{eigen})
one has the determinant
\begin{equation}
\left| R^a_{\  b} - \lambda \delta^a_b \right| \,, \label{PC}
\end{equation}
which is a polynomial of degree five in $\lambda$,
called the {\em characteristic polynomial} of $R^a_{\ b}$. The
fundamental theorem of algebra~\cite{Cajori1969} ensures that, 
over the complex field, it can be always factorized as
\begin{equation} \label{PoliCarac}
(\lambda - \lambda_1)^{d_1} \, (\lambda - \lambda_2)^{d_2} \,\,
\cdots \,\, (\lambda - \lambda_r)^{d_r} \,,
\end{equation} 
where $\lambda_i$  ($i=1,2,\, \cdots \, ,r$) 
are the distinct roots of the polynomial (eigenvalues), and 
$d_i$                     
the corresponding degeneracies. 
To indicate the characteristic polynomial we shall 
introduce a new list \{$d_1\, d_2\, \cdots \,d_r$\} of
eigenvalues degeneracies, hereafter referred to as the
{\em type of the characteristic polynomial.}

The minimal polynomial can be introduced as follows.
Let $P$ be a monic matrix polynomial of degree $n$ in $R^a_{\ b}$, i.e., 
\begin{equation}
P = R^n + c_{n-1} \, R^{n-1} + c_{n-2} \, R^{n-2} + 
\, \cdots \, + c_1\, R + c_0\,\delta \,,
\end{equation}
where $\delta$ is the identity matrix and $c_n$ 
are, in general, complex numbers.
The polynomial $P$ is said to be the {\em minimal polynomial} 
of $R$ if it is the polynomial of lowest degree in $R$ such 
that $P=0$. It can be shown~\cite{Wilkinson1965}
that the minimal (monic) polynomial is unique and can be 
factorized as
\begin{equation} \label{PoliMin}
(R - \lambda_1 \delta)^{m_1} \,
(R - \lambda_2 \delta)^{m_2} \, \cdots \,,
(R - \lambda_r \delta)^{m_r} \,,
\end{equation}
where $m_i$  is the dimension of the       
Jordan block of {\em highest dimension\/} for each eigenvalue
$\lambda_1,\ \lambda_2,\, \cdots \,,\lambda_r,$ respectively. We shall
denote the minimal polynomial through a third list $\| m_1\, m_2\, \cdots\, 
m_r\,\|$, hereafter referred to as the {\em type of the minimal polynomial.} 

We can work out now the characteristic and minimal polynomials for
each Segre type in 5-D. The power $d_i$ of the term corresponding to each 
eigenvalue $\lambda_i$ in the characteristic polynomial is the sum of the 
dimensions of the Jordan blocks with the same eigenvalue $\lambda_i$,
whereas in the minimal polynomial the power $m_i$ is the dimension of the 
Jordan block of highest dimension with that eigenvalue. Thus, for example, 
the Segre types [(1,111)1], [(31)1] and [(211)1] have the same type for
the characteristic polynomial, namely type \{41\}, while their 
corresponding minimal polynomials are, respectively, of types  
$\|11\|$, $\|31\|$ and $\|21\|$. On the other hand, the Segre 
types [3(11)] and [(31)1] have the type $\|31\|$ for the minimal 
polynomial, while the associated characteristic polynomials are, 
respectively, of types \{32\} and \{41\}. We also remark that the
Segre types [2(111)] and [(21)(11)] have the same type for both
polynomials, namely \{32\} and $\|21\|$.

Table~\ref{SegrePCPM} collects together the characteristic (columns - CP) 
and minimal polynomial (rows - MP) types corresponding to the possible 
Segre types of a symmetric two-tensor in 5-D Lorentzian spaces.
It should be noticed that the characteristic polynomial for the 
complex Segre types [$z\bar{z}\,$111], [$z\bar{z}\,$1(11)] and 
[$z\bar{z}\,$(111)] have been denoted, respectively,  
by \{$z\bar{z}$111\}, \{21$z\bar{z}$\} and \{3$z\bar{z}$\}.

\begin{table} {\scriptsize
\begin{center} 
\begin{tabular}{||c|| l| l| l| l| l| l| l| l| l| l||} 
\hline\hline 
CP $\rightarrow$  
           &\{11111\}  &\{$z\bar{z}$111\} & \{2111\}   &\{21$z\bar{z}$\}   &
 \{221\}   & \{311\}   &\{3$z\bar{z}$\}   & \{32\}     & \{41\}            
&  \{5\}    \\
\cline{1-1}
MP$\ \downarrow$ &&&&&&&&&& \\ \hline\hline 
$\|11111\|$ & [1,1111] & [$z\bar{z}$111]  &            &                   &
            &          &                  &            &               
& \\ \hline 
$\|2111\|$  &          &                 & [2111]     &                   &
            &          &                 &            &               
& \\ \hline 
$\|311\|$   &          &                 &            &                   &
            & [311]    &                 &            &                   
& \\ \hline 
$\|1111\|$  &          &                 & [1,11(11)] & [$z\bar{z}$1(11)] &
            &          &                 &            &        
&   \\
            &          &                 & [(1,1)111] &                   &
            &          &                 &            &         
& \\ \hline 
$\|211\|$   &           &                &            &                   & 
[21(11)]   & [(21)11]  &                &            &        
& \\ \hline
$\|31\|$   &            &                &            &                   & 
           &            &                & [3(11)]    &      [(31)1] 
&  \\ \hline 
$\|111\|$  &            &                &            &                   & 
[(1,1)1(11)]&[(1,11)11]&[$z\bar{z}$(111)]&            &      
&  \\
           &             &               &            &                   &
[1,(11)(11)] &[1,1(111)] &               &            & 
&  \\ \hline   
$\|21\|$   &             &               &            &                   &
           &             &               & [2(111)]   &     [(211)1]
& \\ 
            &            &               &            &                   &
            &            &               & [(21)(11)] &  
& \\ \hline 
$\|3\|$     &            &               &            &                   &
            &            &               &            &      
&  [(311)] \\ \hline 
$\|11\|$    &            &               &            &                   &
            &            &               & [(1,11)(11)] & [(1,111)1]        
& \\ 
            &            &               &            &                   &
            &            &               & [(1,1)(111)] & [1,(1111)]
& \\ \hline 
$\|2\|$     &            &               &            &                   &
            &            &               &            &  
& [(2111)] \\ \hline
$\|1\|$     &            &               &            &                   &
            &            &               &            &  
& [(1,1111)] \\ 
\hline\hline
\end{tabular} \end{center}  }
\caption[]{Characteristic (columns - CP) and minimal (rows - MP) polynomial 
types corresponding to the Segre types of $R^a_{\ b}$ in 5-D Lorentzian
spaces.} 
\label{SegrePCPM}
\end{table}

\section{Limiting Diagram for Segre Types in 5-D} 
\label{LimitDiag}	       
\setcounter{equation}{0}

In the study of limits of space-times it is worth noticing that
there are some properties that are inherited by all limits of 
a family of space-times~\cite{Geroch1969}. These properties 
are called hereditary. Thus, for example, a hereditary property 
devised by Geroch can be stated as follows:

\begin{equation} \label{PropH0}
\begin{minipage}[t]{60ex}
{\bf Hereditary property:}\\ 
Let $T$ be a tensor or scalar field built from the metric and 
its derivatives. If $T$ is zero for all members of a family of 
space-times, it is zero for all limits of this family.
\end{minipage} \vspace{3mm}
\end{equation}

Two corollaries of this property are that limits of 
conformally flat space-times are conformally flat,
and that limits of Ricci flat space-times are also 
vacuum solutions in GR.

In general, the algebraic type of the Weyl tensor is not a 
hereditary property under limiting processes. Nevertheless,
to be at least as specialized as the types in the Penrose specialization 
diagram for the Petrov classification is~\cite{Geroch1969}.

Similarly, although the Segre type of the Ricci tensor is
not in general preserved under limiting processes, there exits a
limiting diagram for the Segre types in GR, which has been
recently discussed~\cite{PaivaReboucasHallMacCallum1996}.

In this section, we shall discuss limiting diagrams for 
both the characteristic and minimal polynomial types, and combine 
them to determine a limiting diagram for the Segre types 
of $R^a_{\ b}$ in 5-D Lorentzian spaces. 

Clearly the characteristic~(\ref{PC}) and the minimal~(\ref{PoliMin})
polynomials of $R^a_{\ b}$ as well as the eigenvalues are built from the metric and 
its derivatives~\cite{PaivaReboucasHallMacCallum1996}. Since they
are either scalars or tensors built from the metric and its 
derivatives (hereafter referred to as Geroch scalars and Geroch 
tensors), the hereditary property~(\ref{PropH0}) can be applied 
to them.

A limiting diagram for the types of a five degree characteristic 
polynomial corresponding to the eigenvalue problem~(\ref{eigen})
can now be constructed. We first note that as at each degeneracy one 
Geroch scalar (the difference between two roots of the characteristic 
polynomial) vanishes, by the hereditary property~(\ref{PropH0}), 
under a limiting process, the degeneracy of the characteristic 
polynomial either increases or remains the same.
Besides, the real and imaginary parts of complex roots are also 
Geroch scalars. Therefore, Segre types with real roots 
cannot have as a limit a Segre type with a complex root.
Further, since complex roots can occur only in complex conjugate 
pairs, under a limiting process they either remain complex 
or become a pair of degenerate real roots.
These results can be collected together in the limiting diagram
for the characteristic polynomial shown in figure~\ref{PCEsp}. 
For the sake of simplicity, in the limiting diagrams in this paper, we
do not draw arrows between types whenever a compound limit exists.
Thus, in figure~\ref{PCEsp}, e.g., the limits 
$\{11111\} \rA \{2111\} \rA \{311\} \rA \{32\}$
imply that the limit $\{11111\} \rA \{32\}$ is allowed.

\begin{figure}
\setlength{\unitlength}{3ex}
\begin{center}
\begin{picture}(20,10)(0,-10)

\put(0,0){\makebox(0,1){$\|$11111$\|$}}
\put(0,-2.5){\makebox(0,1){$\|$2111$\|$}}
\put(0,-5){\makebox(0,1){$\|$311$\|$}}

\put(0,-0.1){\vector(0,-1){1.1}}
\put(0,-2.6){\vector(0,-1){1.1}}
\put(1.5,-2){\vector(1,0){1.8}}
\put(1.3,-4.5){\vector(1,0){2}}

\put(5,-2.5){\makebox(0,1){$\|$1111$\|$}}
\put(5,-5){\makebox(0,1){$\|$211$\|$}}
\put(5,-7.5){\makebox(0,1){$\|$31$\|$}}

\put(5,-2.6){\vector(0,-1){1.1}}
\put(5,-5.1){\vector(0,-1){1.1}}
\put(6.5,-4.5){\vector(1,0){2}}
\put(6.3,-7){\vector(1,0){2.2}}

\put(10,-5){\makebox(0,1){$\|$111$\|$}}
\put(10,-7.5){\makebox(0,1){$\|$21$\|$}}
\put(10,-10){\makebox(0,1){$\|$31$\|$}}

\put(10,-5.1){\vector(0,-1){1.1}}
\put(10,-7.6){\vector(0,-1){1.1}}
\put(11.1,-7){\vector(1,0){2.5}}
\put(11.1,-9.5){\vector(1,0){2.5}}

\put(15,-7.5){\makebox(0,1){$\|$11$\|$}}
\put(15,-10){\makebox(0,1){$\|$2$\|$}}

\put(15,-7.6){\vector(0,-1){1.1}}
\put(16,-9.5){\vector(1,0){2.8}}

\put(20,-10){\makebox(0,1){$\|$1$\|$}}
\end{picture}
\end{center}
\caption{Limiting diagram for the minimal polynomial in 5-D Lorentzian
spaces. The types $\|221\|$, $\|32\|$, $\|22\|$, $\|4\|$ and $\|5\|$ are
not shown since they do not correspond to any Segre type (see table
\protect\ref{SegrePCPM}).}
\label{PMEsp} 
\end{figure}
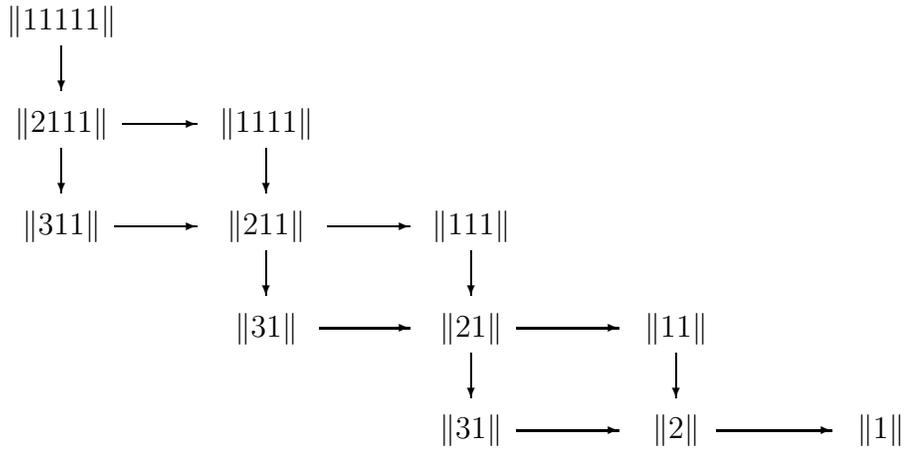

A limiting diagram for the types of a five degree minimal
polynomial of the Ricci tensor can be constructed as follows.
According to the hereditary property (\ref{PropH0}), the minimal
polynomial of a family of space-times is zero for all limits of this
family. Therefore, under limiting processes the degree of the minimal
polynomial either decreases or remains the same. Besides, from the
limiting diagram for the characteristic polynomial in
figure~\ref{PCEsp} we notice that also the number $r$ of distinct 
eigenvalues either decreases or remains the same.
Taking into account these properties we can work 
out the limiting diagram for the minimal polynomial shown 
in figure~\ref{PMEsp}, where the columns correspond to the same 
degree $m_1 + m_2 + \, \cdots \, + m_r$ of the minimal polynomial, 
and the rows correspond to the same number $r$ of distinct  eigenvalues. 
It should be noticed that although we could also have distinguished 
complex from real roots in the minimal polynomial diagram, for our purpose 
in this paper it can be verified that no useful information would arise.

\begin{figure}
\setlength{\unitlength}{3ex}
\begin{center}
\begin{picture}(11,10.5)(0,-10.5)

\put(0,0){\makebox(0,1){\{11111\}}}
\put(0,-3){\makebox(0,1){\{2111\}}}
\put(0,-5.5){\makebox(0,1){\{311\}}}
\put(0,-8){\makebox(0,1){\{41\}}}
\put(0,-10.5){\makebox(0,1){\{5\}}}

\put(0,-0.1){\vector(0,-1){1.6}}
\put(0,-3.1){\vector(0,-1){1.3}}
\put(0,-5.6){\vector(0,-1){1.3}}
\put(0,-8.1){\vector(0,-1){1.3}}

\put(0.2,-3.1){\vector(4,-1){4.5}}
\put(0.2,-5.6){\vector(4,-1){4.5}}

\put(5.5,-5.5){\makebox(0,1){\{221\}}}
\put(5.5,-8){\makebox(0,1){\{32\}}}

\put(5.5,-5.6){\vector(0,-1){1.3}}

\put(5.5,-5.6){\vector(-3,-1){4.5}}
\put(5.5,-8.1){\vector(-3,-1){4.8}}

\put(10,0){\makebox(0,1){\{$z\bar{z}$111\}}}
\put(10,-3){\makebox(0,1){\{21$z\bar{z}$\}}}
\put(10,-5.5){\makebox(0,1){\{3$z\bar{z}$\}}}

\put(10,-0.1){\vector(0,-1){1.6}}
\put(10,-3.1){\vector(0,-1){1.3}}

\put(10,-0.1){\vector(-4,-1){8.2}}
\put(10,-3.1){\vector(-3,-1){4.0}}
\put(10,-5.6){\vector(-3,-1){4.0}}

\end{picture}
\end{center}
\caption{Limiting diagram for the characteristic polynomial in 5-D
Lorentzian spaces.}
\label{PCEsp} 
\end{figure}
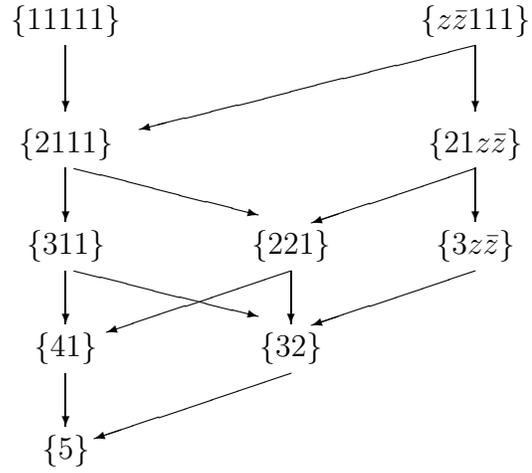

\begin{figure}
\setlength{\unitlength}{3ex}
\begin{center}
\begin{picture}(28,13)(0,-13)

\put(0,0){\framebox(4,2){\shortstack{[11111] \\ {[}$z\bar{z}$111]}}}
\put(0,-2.5){\framebox(4,1){[2111]}}
\put(0,-5){\framebox(4,1){[311]}}

\put(2,-0.1){\vector(0,-1){1}}
\put(2,-2.6){\vector(0,-1){1}}
\put(4.3,-2){\vector(1,0){1}}
\put(4.3,-4.5){\vector(1,0){1}}

\put(6,-2.5){\framebox(4,2){\shortstack{[111(11)] \\ {[}$z\bar{z}$1(11)]}}}
\put(6,-6){\framebox(4,2){\shortstack{[21(11)] \\ {[}(21)11]}}}
\put(6,-9.5){\framebox(4,2){\shortstack{[3(11)] \\ {[}(31)1]}}}

\put(8,-2.6){\vector(0,-1){1}}
\put(8,-6.1){\vector(0,-1){1}}
\put(10.3,-5){\vector(1,0){1}}
\put(10.3,-8.5){\vector(1,0){1}}

\put(12,-6){\framebox(4,3){\shortstack
{[(11)1(11)] \\ {[}(111)11] \\ {[}$z\bar{z}$(111)]}}}
\put(12,-10.5){\framebox(4,3){\shortstack
{[2(111)] \\ {[}(21)(11)] \\ {[}(211)1)]}}}
\put(12,-13){\framebox(4,1){[(311)]}}

\put(14,-6.1){\vector(0,-1){1}}
\put(14,-10.6){\vector(0,-1){1}}
\put(16.3,-8.5){\vector(1,0){1}}
\put(16.3,-12.5){\vector(1,0){1}}

\put(18,-9.5){\framebox(4,2){\shortstack{[(111)(11)] \\ {[}(1111)1]}}}
\put(18,-13){\framebox(4,1){[(2111)]}}

\put(20,-9.6){\vector(0,-1){2}}
\put(22.3,-12.5){\vector(1,0){1}}

\put(24,-13){\framebox(4,1){[(11111)]}}

\end{picture}
\end{center}
\caption{Diagram for the limits of Segre types of $R^a_{\ b}$ in 5-D 
Lorentzian spaces according to the type of the minimal polynomial.}
\label{SegreEspPM} 
\end{figure}
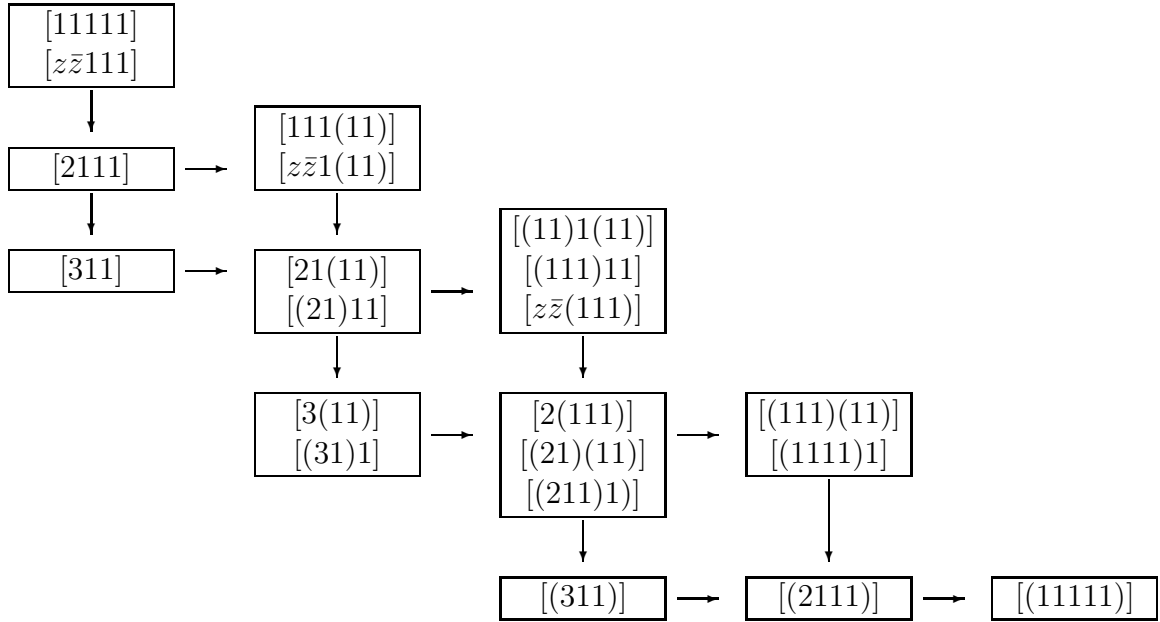

{}From the limiting diagrams for the characteristic and minimal
polynomials in figures~\ref{PCEsp} and~\ref{PMEsp}, we substitute 
for each type of the 
characteristic and minimal polynomials the corresponding Segre types
taken from table~\ref{SegrePCPM}. This gives rise to the two limiting diagrams
given in figures~\ref{SegreEspPC} and \ref{SegreEspPM}, respectively. We have
not taken into account the character of the eigenvectors. Thus, for
example, we represent the Segre types [(1,1)111] and [1,11(11)] by the
set-type [(11)111] and the Segre types [(1,11)(11)] and [(1,1)(111)] by
the set-type [(111)(11)]. The set-types [(11)1(11)] and [(111)11] are
similarly introduced.

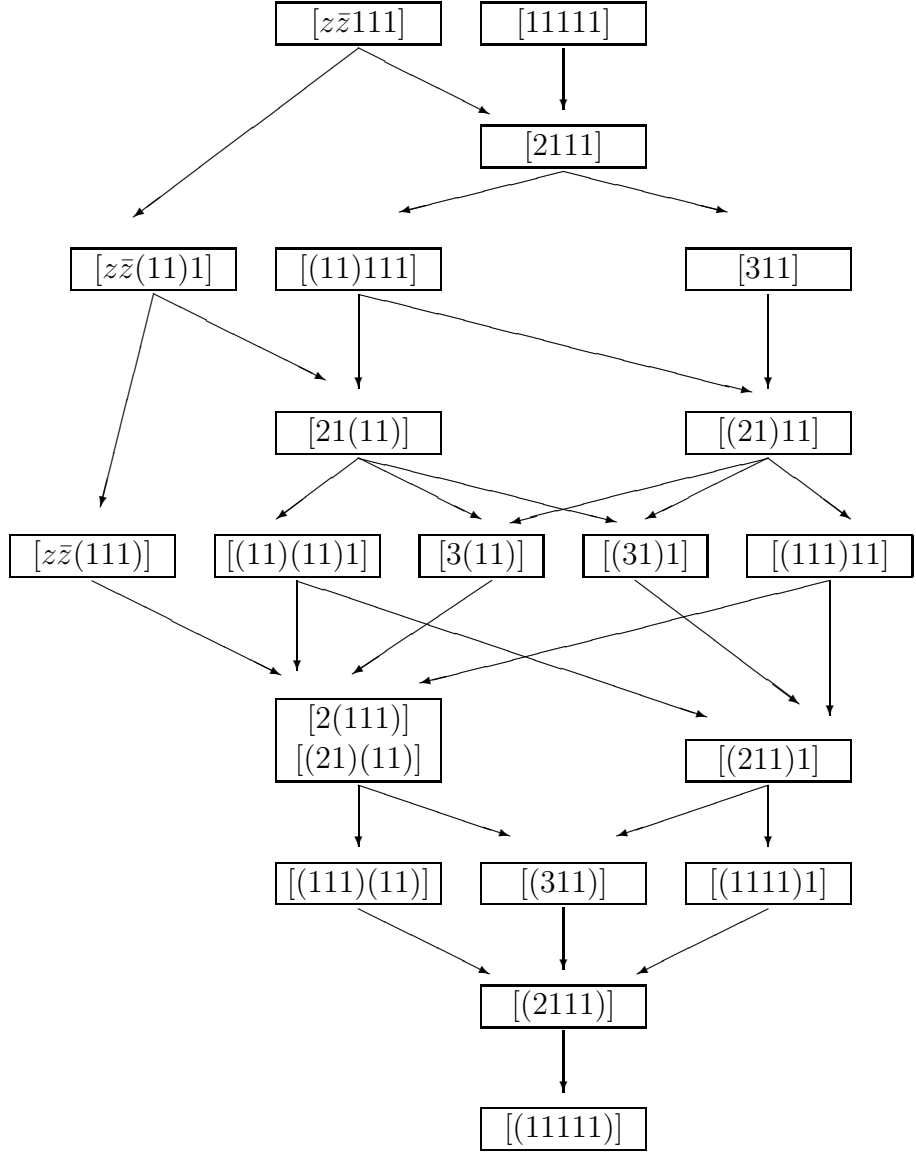
\begin{figure}
\setlength{\unitlength}{3ex}
\begin{center}
\begin{picture}(20,27)(-9.5,-27)

\put(-5,0){\framebox(4,1){[$z\bar{z}$111]}}
\put(0,0){\framebox(4,1){[11111]}}

\put(-3,-0.1){\vector(-4,-3){5.5}}
\put(-3,-0.1){\vector(2,-1){3.2}}
\put(2,-0.1){\vector(0,-1){1.5}}

\put(0,-3){\framebox(4,1){[2111]}}

\put(2,-3.1){\vector(-4,-1){4}}
\put(2,-3.1){\vector(4,-1){4}}

\put(-10,-6){\framebox(4,1){[$z\bar{z}$(11)1]}}
\put(-5,-6){\framebox(4,1){[(11)111]}}
\put(5,-6){\framebox(4,1){[311]}}

\put(-8,-6.1){\vector(-1,-4){1.3}}
\put(-8,-6.1){\vector(2,-1){4.2}}

\put(-3,-6.1){\vector(0,-1){2.3}}
\put(-3,-6.1){\vector(4,-1){9.6}}

\put(7,-6.1){\vector(0,-1){2.3}}

\put(-5,-10){\framebox(4,1){[21(11)]}}
\put(5,-10){\framebox(4,1){[(21)11]}}

\put(-3,-10.1){\vector(-4,-3){2}}
\put(-3,-10.1){\vector(2,-1){3}}
\put(-3,-10.1){\vector(4,-1){6.3}}

\put(7,-10.1){\vector(4,-3){2}}
\put(7,-10.1){\vector(-2,-1){3}}
\put(7,-10.1){\vector(-4,-1){6.3}}

\put(-11.5,-13){\framebox(4,1){[$z\bar{z}$(111)]}}
\put(-6.5,-13){\framebox(4,1){[(11)(11)1]}}
\put(-1.5,-13){\framebox(3,1){[3(11)]}}
\put(2.5,-13){\framebox(3,1){[(31)1]}}
\put(6.5,-13){\framebox(4,1){[(111)11]}}

\put(-9.5,-13.1){\vector(2,-1){4.6}}

\put(-4.5,-13.1){\vector(0,-1){2.2}}
\put(-4.5,-13.1){\vector(3,-1){10}}

\put(0.25,-13.1){\vector(-3,-2){3.4}}

\put(3.75,-13.1){\vector(4,-3){4}}

\put(8.5,-13.1){\vector(-4,-1){10}}
\put(8.5,-13.1){\vector(0,-1){3.3}}

\put(-5,-18){\framebox(4,2){\shortstack{[2(111)] \\ {[}(21)(11)]}}}
\put(5,-18){\framebox(4,1){[(211)1]}}

\put(-3,-18.1){\vector(0,-1){1.5}}
\put(-3,-18.1){\vector(3,-1){3.7}}
  
\put(7,-18.1){\vector(-3,-1){3.7}}
\put(7,-18.1){\vector(0,-1){1.5}}

\put(-5,-21){\framebox(4,1){[(111)(11)]}}
\put(0,-21){\framebox(4,1){[(311)]}}
\put(5,-21){\framebox(4,1){[(1111)1]}}

\put(-3,-21.1){\vector(2,-1){3.2}}
\put(2,-21.1){\vector(0,-1){1.5}}
\put(7,-21.1){\vector(-2,-1){3.2}}

\put(0,-24){\framebox(4,1){[(2111)]}}

\put(2,-24.1){\vector(0,-1){1.5}}

\put(0,-27){\framebox(4,1){[(11111)]}}

\end{picture}
\end{center}
\caption{Diagram for the limits of the Segre types of $R^a_{\ b}$ in 5-D
Lorentzian spaces.}
\label{SegreEsp1} 
\end{figure}

We can now collect together the information of the limiting diagrams of
figures~\ref{SegreEspPC} and \ref{SegreEspPM}, to finally draw a
limiting diagram for the Segre types in 5-D, shown in
figure~\ref{SegreEsp1}. Thus, for example, starting from the limiting 
diagram for the minimal polynomial (figure~\ref{SegreEspPM}) one finds
that the Segre type [2111] may have as its limit the types [311],
[11(111)] and [$z\bar{z}$1(11)]. However, from the diagram for the
characteristic polynomial shown in figure~\ref{PCEsp}
one finds that the Segre type [2111]
cannot have the type [$z\bar{z}$1(11)] as its limit. So,  
we have only two arrows starting from the Segre type [2111],
as it has been drawn in diagram~\ref{SegreEsp1}. Although the other 
arrows in figure~\ref{SegreEsp1} can be similarly determined, we
shall not discuss them here for the sake of brevity.

To close this section we remark that again in the limiting diagram shown 
in figure~\ref{SegreEsp1} the character of the eigenvectors is not 
taken into account. We shall return to this point in the next section.

\begin{figure}
\setlength{\unitlength}{3ex}
\begin{center}
\begin{picture}(17,18)(0,-18)

\put(0,0){\framebox(4,1){[11111]}}
\put(0,-3.5){\framebox(4,2){\shortstack{[2111] \\ {[}111(11)]}}}
\put(0,-8){\framebox(4,3){\shortstack{[311] \\ {[}(21)11] \\ {[}(111)11]}}}
\put(0,-12.5){\framebox(4,3){\shortstack
{[(31)1] \\ {[}(211)1] \\ {[}(1111)1]}}}
\put(0,-18){\framebox(4,3){\shortstack
{[(311)] \\ {[}(2111)] \\ {[}(11111)]}}}

\put(2,-0.1){\vector(0,-1){1}}
\put(2,-3.6){\vector(0,-1){1}}
\put(2,-8.1){\vector(0,-1){1}}
\put(2,-12.6){\vector(0,-1){2}}

\put(2,-3.6){\vector(3,-1){5.8}}
\put(2,-8.1){\vector(4,-1){4.6}}

\put(6,-8){\framebox(4,2){\shortstack{[21(11)] \\ {[}(11)1(11)]}}}
\put(6,-13.5){\framebox(4,4){\shortstack
{[3(11)] \\ {[}2(111)] \\ {[}(21)(11)] \\ {[}(111)(11)]}}}

\put(8,-8.1){\vector(0,-1){1}}

\put(8,-8.1){\vector(-4,-1){4.6}}
\put(8,-13.6){\vector(-4,-1){4.6}}

\put(12,0){\framebox(4,1){[$z\bar{z}$111]}}
\put(12,-3.5){\framebox(4,1){[$z\bar{z}$1(11)]}}
\put(12,-8){\framebox(4,1){[$z\bar{z}$(111)]}}

\put(14,-0.1){\vector(0,-1){2}}
\put(14,-3.6){\vector(0,-1){3}}

\put(14,-0.1){\vector(-4,-1){9.8}}
\put(14,-3.6){\vector(-3,-1){5.8}}
\put(14,-8.1){\vector(-4,-1){4.6}}

\end{picture}
\end{center}
\caption{Diagram for the limits of Segre types of $R^a_{\ b}$ in 5-D
Lorentzian spaces according to the types of the characteristic polynomial.}
\label{SegreEspPC} 
\end{figure}
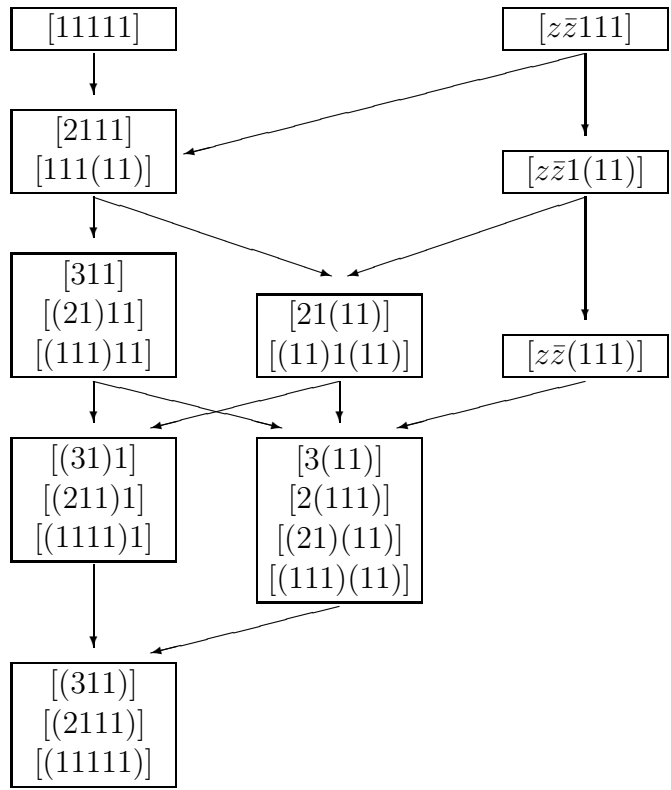

\section{Conclusion} 
\label{Conclusion}
\setcounter{equation}{0}

In this work we have constructed a limiting diagram for
the Segre classification of a second order symmetric 
two-tensor defined on 5-D Lorentzian spaces (figure~\ref{SegreEsp1}). 
To achieve this goal we have essentially used the hereditary
property~(\ref{PropH0}) together with the limiting diagrams 
for the characteristic and minimal polynomial types,
which we have worked out in section~\ref{LimitDiag}.

Improvements of the limiting diagram that we have presented
in this article can still be tackled. A first refinement would
arise by taking into account the character of the eigenvectors.
To take into account the types which differ by the character
of the eigenvectors one has firstly to separate the set-types
into its two members, and check whether one of these members
can have the other as its limits and vice-versa. Secondly,
one needs to find out whether the Segre types which can have as 
limit one set-type can have a limit both members of the set.
Finally, one ought to examine whether the Segre types which can
be a limit of one set-type can be a limit of each member
of the corresponding Segre set-type.
Perhaps most of these checkings can be made simply by extending 
to 5-D space-times the hereditary properties discussed in 
the context of GR~\cite{PaivaReboucasHallMacCallum1996}.
A second refinement of the limiting diagram shown in 
figure~\ref{SegreEsp1} can be made by figuring out 
a criterion for separating the Segre types [2(111)] and 
[(21)(11)], which have the same type for both characteristic
and minimal polynomials. 
A third improvement of the limiting diagram given in
figure~\ref{SegreEsp1} might arise if besides the type of 
the characteristic and minimal polynomials one considers the 
values of their roots. 

The limiting diagrams of the Petrov and the Segre classification 
play a fundamental role in the study of limits of space-times
in general relativity%
~\cite{PaivaReboucasMacCallum1993,PaivaRomero1993,Paiva1993}, as
briefly discussed in the introduction. 
Although the coordinate-free technique for finding out limits of 
space-times in GR~\cite{PaivaReboucasMacCallum1993}
have not yet been extended to 5-D space-times, the limiting 
diagram studied in the present work will certainly be applicable to 
any coordinate-free approach to possible limits of non-vacuum 
space-times in five dimensions.

\vspace{4mm}
\section*{Acknowledgment}
The authors gratefully acknowledge financial assistance from CNPq.


\end{document}